\documentclass[aip,apl,reprint,twocolumn,noeprint]{revtex4-2}

\usepackage[dvipdfmx]{graphicx}

\usepackage[version=3]{mhchem}

\usepackage{braket}

\usepackage{natbib}

\usepackage{here}

\usepackage{dcolumn}

\usepackage{color}

\usepackage{bm}

\begin{document}
\title{Optically-induced magnetization switching in NiCo$_{2}$O$_{4}$ thin films using ultrafast lasers}  
\author{Ryunosuke Takahashi}  \email{ri20a017@stkt.u-hyogo.ac.jp}
\affiliation{Department of Material Science, Graduate School of Science, University of Hyogo, Ako, Hyogo 678-1297, Japan}
\author{Takuo Ohkochi}
\affiliation{Japan Synchrotron Radiation Research Institute, Sayo, Hyogo, 679-5198, Japan}
\affiliation{RIKEN SPring-8 Center, Kouto, Sayo, Hyogo, 679-5148, Japan}
\author{Daisuke Kan}
\author{Yuichi Shimakawa}
\affiliation{Institute for Chemical Research, Kyoto University, Uji, Kyoto 611-0011, Japan}
\author{Hiroki Wadati}
\affiliation{Department of Material Science, Graduate School of Science, University of Hyogo, Ako, Hyogo 678-1297, Japan}
\affiliation{Institute of Laser Engineering, Osaka University, Suita, Osaka 565-0871, Japan}

\begin{abstract}
Recently, all-optical magnetization control has been garnering considerable attention in realizing next-generation ultrafast magnetic information devices. Here, employing a magneto-optical Kerr effect (MOKE) microscope, we observed the laser-induced magnetization switching of ferrimagnetic oxide NiCo$_{2}$O$_{4}$ (NCO) epitaxial thin films with perpendicular magnetic anisotropy, where the sample was pumped at 1030-nm laser pulses, and magnetic domain images were acquired via the MOKE microscope with a white light emitting diode. Laser pulses irradiated an NCO thin film at various temperatures from 300 K to 400 K while altering the parameters of pulse interval, fluence, and the number of pulses with the absence of the external magnetic field. We observed accumulative all-optical switching at 380 K and above. Our observation of oxide NCO thin films facilitates the realization of chemically stable magnetization switching using ultrafast lasers, and without applying a magnetic field.  

\end{abstract}

\maketitle
Ultrafast magnetization switching by pulsed lasers has been actively studied recently owing to its potential as a next-generation magnetic recording device. Laser-induced demagnetization and magnetization switching were pioneered by the detection of ultrafast demagnetization of less than 1 ps in Ni foil by Beaurepaire {\sl et al} \cite{Beaurepaire1996}. More comprehensive studies have revealed that when a magnetic material is irradiated with laser pulses, the magnetization changes on a femtosecond time scale \cite{Beaurepaire1998,Hohlfeld2001,Koopmans2003}. Subsequently, to elucidate the mechanisms of ultrafast demagnetization and magnetization switching, laser irradiated magnetic domains were also observed via microscopic measurements \cite{Stanciu2007prl,Ostler2012}. Currently, studies on ultrafast laser effects have been widely conducted, including laser-induced magnetization switching observed for each element and theoretical study \cite{Kachel2009,Radu2011,Takubo2017,Yamamoto2020,Koopmans2010,Dornes2019}; however, the mechanism for this phenomenon remains a debate. 

A phenomenon where magnetization is solely switched by laser irradiation without applying a magnetic field is known as all-optical switching (AOS), and has been observed in magnetic thin films such as GdFeCo ferrimagnetic alloys, TbFeCo ferrimagnetic alloys, and Co/Pt ferromagnetic multilayer films. \cite{Stanciu2007,Vahaplar2009,Ostler2012,Ohkochi2012, Lambert2014,TbFeCo2015,CoPtswitching2017,Parlak2018}. There are two types of AOS for magnetization: all-optical  {\sl helicity-dependent} switching (AO-HDS) and all-optical {\sl helicity-independent} switching (AO-HIS). AO-HDS was experimentally observed in ferromagnetic materials such as Co/Pt multilayer films\cite{CoPtswitching2017,Parlak2018}. AO-HIS can be observed in ferrimagnetic materials owing to the existence of two spin sublattices, where either circularly polarized light can excite the electron spin. Regarding GdFeCo ferrimagnetic alloys, single-shot AO-HIS can be observed \cite{Stanciu2007,Ostler2012,Iihama2018}. AO-HDS was also observed in GdFeCo ferrimagnetic alloys under certain conditions) in which the inverse magneto-optical effect is well featured by clear AO-HIS. The perfect AO-HIS is characteristic of the ferrimagnetic system's response, caused by the difference in the onset time of demagnetization between transition-metal and rare-earth sub-lattices \cite{Stanciu2007}. 
Parlak {\sl et al}. investigated AO-HDS in Co/Pt multilayer films using a magneto-optical Kerr effect (MOKE) microscope. When the Co/Pt multilayer film was irradiated with 800-nm laser pulses, they observed accumulative AO-HDS with an all-optical domain formation (AODF) within the laser-induced demagnetized area owing to the effect of thermalization. The AO-HDS in the Co/Pt system has been observed as a ring-shaped uniformly-switched domain, with a randomly distributed domain at the central irradiation spot; the AO-HDS is realized at the significantly limited peripheral area where the local heat load given by a Gaussian beam is at the AOS threshold. \cite{Parlak2018}. Presently, the search for AOS in oxides and rare-metal-free materials is increasingly important, considering its future device applications.
\begin{figure*}[t]
\centering
\includegraphics[width = 12cm]{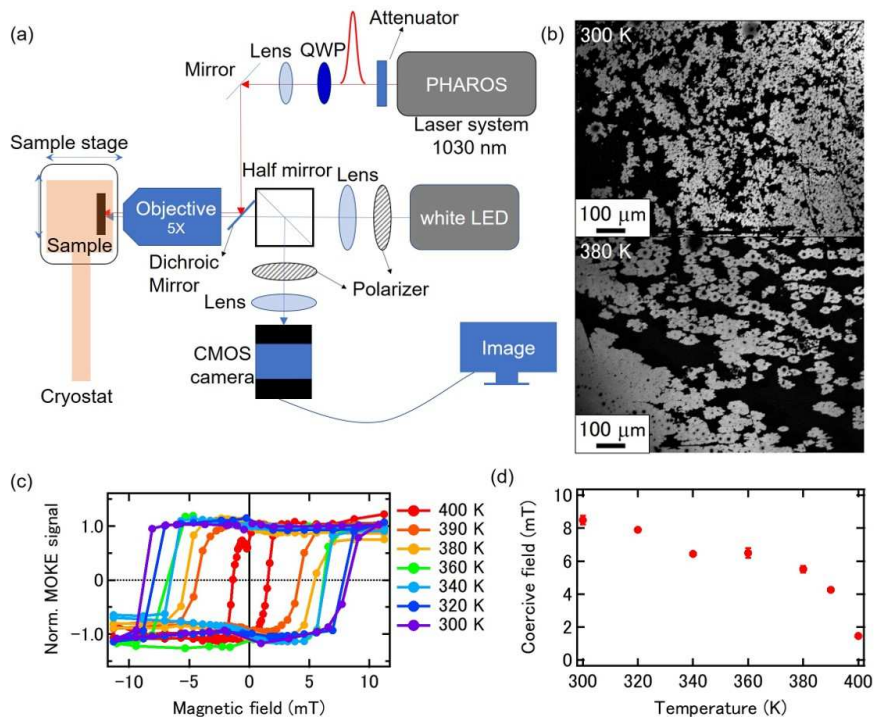}
\caption{(a) Diagram of MOKE microscope setup. (b) Magnetic domain images of the NCO thin film at temperatures of 300 K and 380 K observed via MOKE microscopy measurements. (c) Magnetic hysteresis loops of the NCO thin film as a function of the magnetic field out of plane at temperatures between 300 K and 400 K. (d) Temperature dependence of the coercive magnetic field. The scale bars are 100 $\mu$m}
\label{fig1}
\end{figure*}
NiCo$_{\bm{2}}$O$_{\bm{4}}$ (NCO) with an inverse-spinel-type crystal structure is a ferrimagnetic oxide whose Curie temperature ($T_{C}$) is higher than 400 K \cite{Knop1968}. Co occupies the $T_{d}$ and $O_{h}$ sites, while Ni occupies the $O_{h}$ site. The valence states are close to Co$^{2+}$ ($O_{h}$ site), Co$^{3+}$ ($T_{d}$ site), and Ni$^{2+\delta}$ ($O_{h}$ site), as determined via X-ray absorption spectroscopy and XMCD measurement \cite{Kan2020prb}. The spin directions of $T_{d}$-site Co and $O_{h}$-site Ni are ferrimagnetically coupled \cite{Bitla2015,Kan2020prb}, and their saturated magnetizations have been determined to be approximately 2 $\mu_{B}$ \cite{Kan2020}. NCO thin films exhibit the property of perpendicular magnetic anisotropy (PMA), which easily magnetizes in the out-of-plane direction. PMA is noteworthy because of its potential for higher density magnetic recording, which enables the realization of smaller magnetic devices. We previously reported that NCO thin films realize the ultrafast demagnetization of approximately 0.4 ps at 300 K, thus indicating that NCO films have the spin polarization as large as approximately 0.7 \cite{Takahashi2021}. 
Owing to these suitable features, NCO thin films have been garnering attention as spintronic materials with potential applications in next-generation high-density non-volatile memory devices that are stable at variable temperatures \cite{Chen2019AM,Shen2020,Tsujie2020apl,MellingerPhysRevB.101}.

In this study, to observe the AOS in the NCO thin film, we performed MOKE microscopy measurements of laser-irradiated magnetic domains at various temperatures between 300 K and 400 K. We revealed that the accumulative AOS ring was observed at 380 K and above, in contrast to the previously known GdFeCo alloys and Co/Pt multilayer films where AOS emerged at 300 K. Hence, accumulative AOS was realized in the oxide NCO thin film by increasing the temperature.

Epitaxial 30-nm thick NCO thin films were grown on MgAl$_{\bm{2}}$O$_{\bm{4}}$(100) substrates using pulsed laser deposition. During the film deposition, the substrate temperature and oxygen pressure were 315$^\circ$C and 100 mTorr, respectively. Details of the NCO sample preparation are described in Refs. \cite{Kan2020,Shen2020}.

\begin{figure*}[t]
\centering
\includegraphics[width = 12cm]{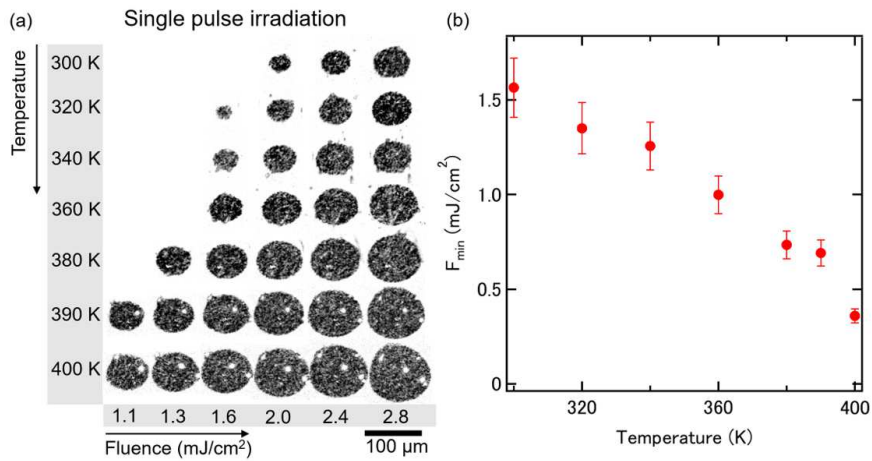}
\caption{(a) Temperature and fluence dependent laser-induced magnetic domains of the NCO thin film when irradiating 1 pulse. (b) Temperature dependent threshold of laser fluence. The scale bar in (a) is 100 $\mu$m}
\label{fig3}
\end{figure*}
To observe the laser irradiated effect of the NCO thin film, we combined a femtosecond laser with a MOKE microscope. Figure~\ref{fig1} (a) presents the experimental setup using white light emitting diode (LED) as the light source with the K\"{o}hler illumination method. The sample was set in a cryostat equipped with a heater to change the ambient temperature of the sample from 300 K to 400 K. The sample was irradiated with linearly horizontal laser pulses using Yb:KGW laser Pharos ($\lambda$ = 1030 nm ($h\nu$ = 1.2 eV), 1 kHz, FWHM $\sim$ 200 fs) as the light source. This pump laser beam was focused by a lens with a focal length of 200 mm, and the diameter of the laser spot was approximately 180 $\mu$m in FWHM at the sample position. A quarter-wave plate (QWP) was introduced to transform the horizontally linear polarization ($\pi$) of the laser into circular polarization. The “left ($\sigma^-$)” or “right ($\sigma^+$)” circular polarization of the laser was adjusted by rotating the QWP by ±45$^{\circ}$ with respect to the plane of linear polarization. An attenuator was introduced to arrange the power of the laser. The NCO thin film was uniformly magnetized by a permanent magnet before irradiation with the laser pulses. Magnetic domain images were observed with a complementary metal oxide semiconductor (CMOS) camera using an Olympus objective lens 5$\times$, and the field of view was 1000 $\times$ 750 $\mu$m with the angle set between two polarizers as 45$^\circ$. Figure~\ref{fig1} (b) presents the magnetic domain images of the NCO thin film by the MOKE microscope at 300 K and 380 K at 0 mT. It can be observed that the domain size of 380 K is larger than that of 300 K, because the magnetization decreases as the sample approaches $T_{C}$. Before irradiating the laser pulses, we investigated the magnetic property of the NCO thin film. Figure~\ref{fig1} (c) illustrates the magnetic hysteresis loops of the NCO thin film obtained by MOKE microscope images. The temperature-dependence of the coercive field $H_{C}$ estimated from Fig.~\ref{fig1} (c) is plotted in Fig.~\ref{fig1} (d). This indicates that the H$_{C}$ value, which is approximately 8 mT at 300 K, abruptly drops above 380 K.
 
\begin{figure}[h]
\begin{center}
\includegraphics[width = \linewidth]{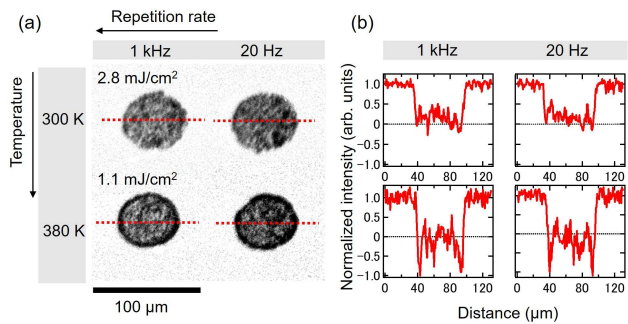}
\caption{(a) Magnetic domain images at 300 K and 380 K after 10$^4$ pulses irradiating with different laser pulse intervals. (b) Contrast values along the dotted lines in panel (a).}
\label{fig4}
\end{center}
\end{figure}

Figure~\ref{fig3} (a) presents the magnetic domain images of the NCO thin film when a single pulse of linearly-polarized laser was irradiated without applying a magnetic field at varying fluences and temperatures. It can be observed that larger a AODF was observed with the same fluence by increasing the temperature. We estimated the minimum fluence $F_{min}$ required for the onset of AODF using the Liu method, as illustrated in Fig.~\ref{fig3} (b) \cite{Liu1982}. This is similar in behavior to the $H_{C}$, decreasing toward 400 K, as shown in Fig.~\ref{fig1} (d).

\begin{figure*}[ht]
\begin{center}
\includegraphics[width = 16cm]{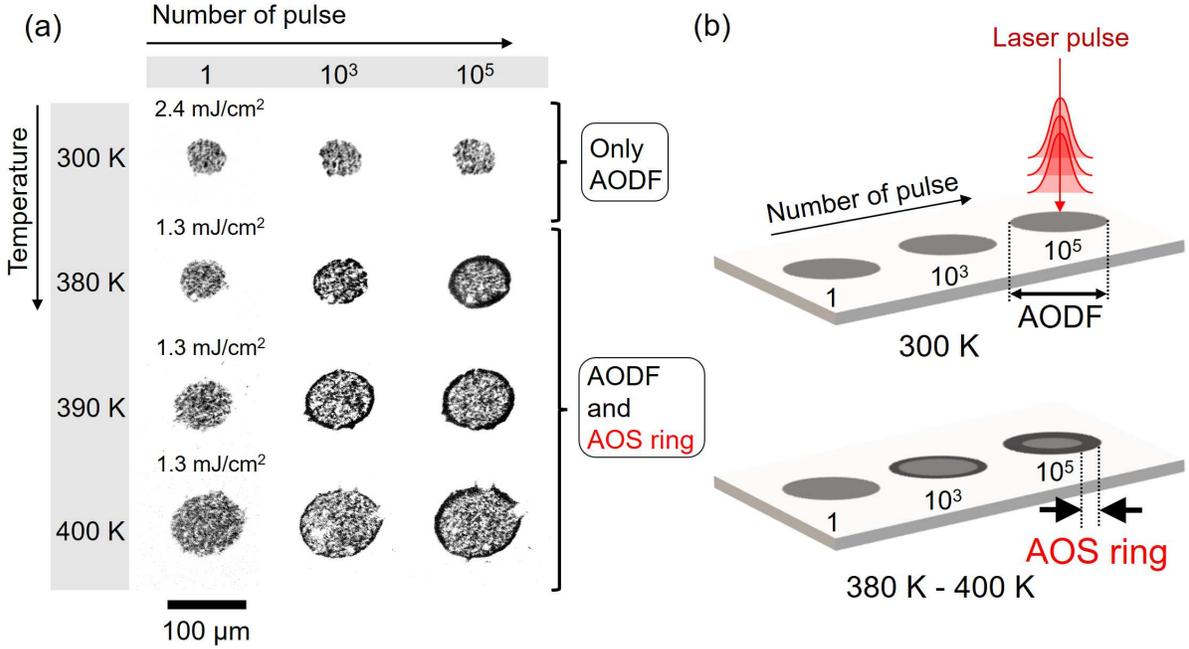}
\caption{(a) Magnetic domain images at 300 K, 380 K, 390 K, and 400 K when irradiated with varying numbers of laser pulses. (b) Schematics of the laser-pulse-accumulation effect in the NCO thin film. The scale bar is 100 $\mu$m}
\label{fig5}
\end{center}
\end{figure*}

To identify AOS via laser-pulse accumulation, we investigated how the AODF and AOS rings depend on laser pulse repetition rate and accumulation by adopting linearly polarized laser pulses. Figure~\ref{fig4} (a) presents magnetic domain images of the NCO thin film after irradiating the identical 10$^4$ pulses , but with different pulse-to-pulse time intervals of 1 and 50 ms corresponding to repetition frequencies of 1 kHz and 20 Hz at 300 K and 380 K, respectively. To compare clearly, we chose the data which show similar AODF size at 1 kHz, namely, 2.8 mJ/cm$^2$ and 1.1 mJ/cm$^2$ for 300 K and 380 K, respectively. Unlike the Co/Pt multilayer case \cite{Parlak2018}, the size of the AODF area is independent of the pulse intervals at both temperatures. This implies that the heat-driven domain wall motion is severely slow owing to the creeping domain wall motion of the NCO thin film with a velocity of approximately 1 $\mu$m/s, as demonstrated in our previous report \cite{Takahashi2021}. Figure~\ref{fig4} (b) presents the contrast values along the dotted lines of the domain images in Fig.~\ref{fig4} (a). We normalized the contrast values of the magnetic domains with up and down spins to $+1$ and $-1$, respectively. At 380 K, it can be observed that the AOS ring was formed on the edge of the AODF. In the AOS area, the contrast value became $-1$, thus indicating magnetization switching. AOS was not observed at 300 K, and the average contrast value in the AODF area was almost 0. This indicates that magnetic domains with up and down spins were randomly mixed.

We also demonstrated pulse-accumulation dependence of linearly-polarized laser-induced magnetization switching effects at 300 K, 380 K, 390 K, and 400 K. The repetition rate of the laser pulse was set to 1 kHz.
As illustrated in Fig.~\ref{fig5} (a), AOS did not emerge with any accumulation of laser pulses at 300 K. AOS can be observed at 380 K and above with the accumulation of laser pulses. Figure~\ref{fig5} (b) presents the schematics of the laser-pulse-accumulation effect of laser-induced magnetization switching in the NCO thin film. At 300 K, where $H_{C}$ is large, AODF was observed without any sign of AOS. At 380 K and above, AOS emerged at the perimeter of AODF irradiating 10$^3$ and 10$^5$ pulses. Furthermore, the AOS area increased by accumulating laser pulses. Hence, AOS emerged in the ferrimagnetic oxide NCO thin film by increasing temperature and accumulating laser pulses.

\begin{figure}[h]
    \centering
    \includegraphics[width=\linewidth]{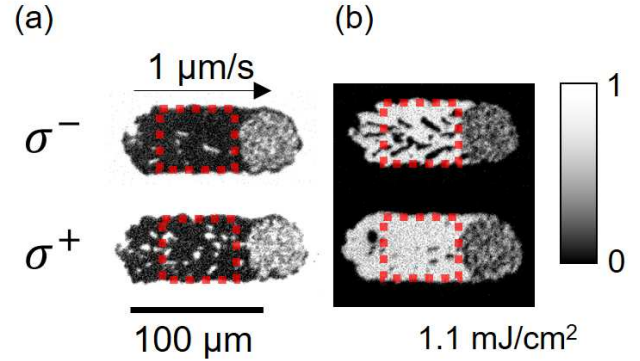}
    \caption{Magnetic domain images of the NCO thin film when the left and right circularly polarized lasers are irradiated into a saturated magnetic domain with up (a) and down (b) spins. The rectangles indicate the parts used to obtain the degree of helicity dependence. (See the text for details).}
    \label{fig6}
\end{figure}

To investigate the helicity dependence of AOS, we irradiated the $\sigma^-$ and $\sigma^+$ circularly polarized lasers into saturated magnetic domains with up (a) and down (b) spins, as illustrated in Fig.~\ref{fig6}. Here, the sample temperature was 380 K and no magnetic field was applied: the condition where the AOS ring was observed in the experiments with linearly polarized pulses presented in Fig.~4. The laser fluence and repetition rate were 1.1 mJ/cm$^{2}$ and 1 kHz, respectively.
The sample was moved at a velocity of 1 $\mu$m/s, and the effect of laser irradiation was observed as a dragging mark from left to right. This dragging method makes it easier to understand the effect of AOS.
By moving the sample at a velocity of 1 $\mu$m/s, approximately 10$^3$ pulses are irradiated at one position of the sample; hence, the area surrounded by the rectangle reflects the effect of AOS.
The results exhibited a weak but non-zero helicity dependence.
To investigate the degree of helicity dependence, we estimated the value of the average contrast $\Delta$$I$ = [$I$($\sigma^-$)-$I$($\sigma^+$)]/[$I$($\sigma^-$)+$I$($\sigma^+$)], where $I$($\sigma^+$) and $I$($\sigma^-$) denote image intensities averaged out over the red rectangular area for $\sigma^+$ and $\sigma^-$ irradiations, respectively. The obtained $\Delta$$I$ value is approximately 0.1 for both the initial up (Fig.~6 (a)) and down (Fig.~6 (b)) domain cases. This result is in stark contrast to the 100\% helicity dependence in reported results of the [Co/Pt]$_3$ film (Fig.~5 (a) in Ref. [U. Parlak et al., Phys. Rev. B, 98, 214443 (2018).]. 

\begin{figure}[h]
\begin{center}
\includegraphics[width = \linewidth]{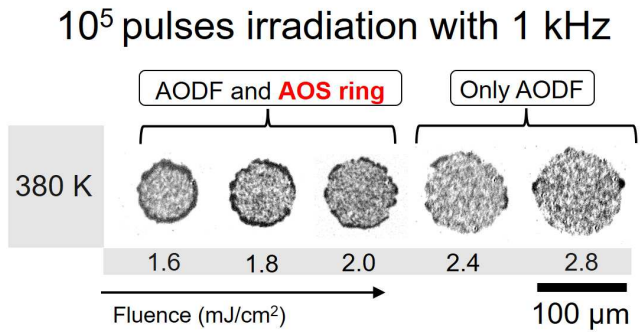}
\caption{Magnetic domain images after irradiating with linearly polarized 10$^5$ pulses at 380 K and several laser fluences. The scale bar is 100 $\mu$m}
\label{fig7}
\end{center}
\end{figure}
Figure~\ref{fig7} presents the fluence dependent magnetic domain images at 380 K. Linearly polarized 10$^5$ pulses with 1-kHz repetition rates irradiated the NCO thin film. It can be observed that both the AOS ring and AODF can be observed when the fluence is less than 2.0 mJ/cm$^2$; however, only AODF was observed when the fluence is 2.4 mJ/cm$^2$ and above.

Now, we discuss the relationship between AOS creation and a stable magnetic domain size. The size of the magnetic domain is determined by the competition between the magnetostatic and exchange interaction energies. The magnetostatic energy is proportional to the square of the magnetization of the sample. Therefore, smaller magnetic domains are expected to be formed  increasing the magnetization. In our proposed NCO thin film, the size of the magnetization was approximately halved as the sample was heated from 300 K to 400 K. At 380 K and above, the size of the stable magnetic domain was so large that AOS ring could exist at the perimeter of AODF. However, at 300 K, the unstable magnetic domains formed by laser irradiation moved toward the spot center owing to the heat-driven domain wall motion. Regarding the laser fluence above 2.4 mJ/cm$^2$, no AOS ring was observed even at 380 K. This is because the size of AODF is large and AOS becomes unstable. This result is consistent with the report in Ref. \cite{Mangin2016}, which suggests that the size of the magnetic domain must be larger than that of the perimeter of AODF for the realization of the AOS ring.

Now, we would like to compare the results of the NCO thin film to the reported results of Co/Pt multilayer films \cite{Parlak2018}. Accordingly, three major differences were observed. First, there was a weak helicity dependence of the laser irradiation effect in the NCO thin film, which is in contrast to the Co/Pt multilayer films \cite{Parlak2018}. Second, accumulative AOS was realized at 380 K and above in the NCO thin films in contrast to Co/Pt multilayer films, where accumulative AOS was realized even at 300 K \cite{Parlak2018}. Third, AODF and AOS areas were not reduced by decreasing the laser repetition rate; however, the areas were reduced because of the heat-driven domain wall motion toward the spot center in the Co/Pt multilayer films \cite{Parlak2018}. This occurs because the domain wall motion is one hundred times slower than that of the Co/Pt multilayer films \cite{Parlak2018}. 
As aforementioned, NCO thin films have an advantage, as AOS can realize by linearly polarized light. Moreover, NCO thin films can be realized magnetization switching regardless of the laser frequency and helicity, which is in contrast to Co/Pt multilayer films. Hence, it is expected that AOS can be stably realized in an oxide thin film.  

In summary, we studied the magnetization switching of NCO thin films under laser irradiation at varying temperatures from 300 K to 400 K.
In particular, we focused on how the results depend on the number of laser pulses, repetition rate, and intensity.
Although AOS was not observed at 300 K, accumulative AOS was observed at 380 K and above.
This result does not depend on the laser repetition rate, thus reflecting the slow domain wall motion of the magnetic domain of the NCO thin film.
The obtained results indicate that the ferrimagnetic oxide thin film with PMA and AOS has the potential to play a vital role in a stable photo-switching device in the future.

This work was supported by the Murata Science Foundation, JSPS KAKENHI under Grant Nos.~19H01816, 19H05816, 19H05823, 19H05824, and 21H01810 KAKENHI and the
MEXT Quantum Leap Flagship Program (MEXT Q-LEAP) under Grant No. JPMXS0118068681. In addition, this study was supported by the Japan Society for the Promotion of Science (JSPS) Core-to-Core Program (A. Advanced Research Networks) and the International Collaborative Research Program of Institute for Chemical Research in Kyoto University, Ministry of Education, Culture, Sports, Science and Technology (MEXT), Japan.

The data that support the findings of this study can be obtained
from the corresponding author upon a reasonable request.

\end{document}